Title:

# Infrared Beam-shaping on Demand via Tailored Geometric Phase Metasurfaces employing the Plasmonic Phase-Change Material In$_3$SbTe$_2$

*Author(s), and Corresponding Author(s)\**

*Lukas Conrads$^{+,*,1,}$, Florian Bontke$^{+,1}$, Andreas Mathwieser$^2$, Paul Buske$^3$, Matthias Wuttig$^1$, Robert Schmitt$^2$, Carlo Holly$^{3,4}$, Thomas Taubner$^{\#,1}$*

+ both authors contributed equally

**Affiliations**

[1] Institute of Physics (IA), RWTH Aachen University, D-52056 Aachen, Germany

[2] Fraunhofer Institute for Production Technology IPT, 52056 Aachen, Germany

[3] Chair for Technology of Optical Systems, RWTH Aachen University, 52056 Aachen, Germany

[4] Fraunhofer Institute for Laser Technology ILT, 52056 Aachen, Germany

\* Email: conrads@physik.rwth-aachen.de

\# Email: taubner@physik.rwth-aachen.de

**Abstract:** (150 / max 150 words)

Conventional optical elements are bulky and limited to specific functionalities, contradicting the increasing demand of miniaturization and multi-functionalities. Optical metasurfaces enable tailoring light-matter interaction at will, especially important for the infrared spectral range which lacks commercially available beam-shaping elements. While the fabrication of those metasurfaces usually requires cumbersome techniques, direct laser writing promises a simple and convenient alternative. Here, we exploit the non-volatile laser-induced insulator-to-metal transition of the plasmonic phase-change material In$_3$SbTe$_2$ (IST) for optical programming of large-area metasurfaces for infrared beam-shaping. We tailor the geometric phase of metasurfaces with rotated crystalline IST rod antennas to achieve beam steering, lensing, and beams carrying orbital angular momenta. Finally, we investigate multi-functional and cascaded metasurfaces exploiting enlarged holography, and design a single metasurface creating two different holograms along the optical axis. Our approach facilitates fabrication of large-area metasurfaces within hours, enabling rapid-prototyping of customized infrared meta-optics for sensing, imaging and quantum information.

**Main:**

Metasurfaces composed of metallic or dielectric nanoantennas of subwavelength size have been widely used to tailor the properties of scattered light such as amplitude, phase, and polarization.[1,2] Numerous antenna geometries with varying sizes and shapes exploiting multiple resonance modes have enabled various functionalities such as beam steering[3], lensing[4], or holography[5]. However, full 2π phase control often requires sophisticated antenna designs with complex shapes and multiple resonance modes. Another promising approach to alter the phase of the scattered light relies on employing geometrically rotated antennas of the same size in combination with circularly polarized light.[6] Accordingly, geometric phase, also called Pancharatnam-Berry phase, metasurfaces facilitate applications such as non-linear optics[7], manipulation of orbital angular momentum[8], and spin-selective deflection[9,10].

The fabrication of these metasurfaces usually relies on cumbersome and complex fabrication techniques involving multiple lithography and etching steps. Direct laser writing of metasurfaces hence offers great potential for speeding up the fabrication and enabling rapid prototyping of various metasurfaces with different functionalities.[11]

Among others[12], promising materials for direct laser writing of metasurfaces are phase-change materials (PCMs).[13] These materials exhibit at least two (meta-) stable phases, the amorphous one and the crystalline one, which differ significantly in their optical and electrical properties.[14,15] The strong refractive index contrast between both phases is attributed to a new bonding mechanism in the crystalline phase, called metavalent.[16–19] This makes PCMs prime candidates for non-volatile dynamic metasurface tuning based on a change in the refractive index.[20,21] In the past years, numerous applications such as active beam steering[9,22], lensing[9,13], and shaping thermal emission[23] have been demonstrated.

Recently, the PCM $In_3SbTe_2$ (IST) has been introduced, which shows a Drude-like behavior with a negative real part of the permittivity in its crystalline phase for the entire infrared spectral range.[24] Therefore, it is classified as a plasmonic PCM enabling direct writing of metallic nanostructures by locally crystallizing the PCM with precise laser pulses. Active resonance tuning can be achieved by modifying the antenna structures themselves and allows for reprogramming once written metasurfaces. Not only tuning antenna resonances by reconfiguring the antenna geometries[25–28], but also active metasurfaces for thermal emissivity shaping[29,30] and tailoring polariton responses[31–33] have been demonstrated.

However, rapid-prototyping of complex IST metasurfaces for real-world applications has not been shown yet. Tailored phase-modulated metasurfaces with the plasmonic PCM IST promise arbitrary beam-shaping in the infrared spectral range, which inherently lacks commercially available devices.

Here, we take advantage of the interplay of circularly polarized light with rotated rod antennas to encode multiple metasurfaces optically written directly into IST. We use a commercial direct laser-writing system for programming large-area metasurfaces and investigate their respective functionalities. First, we demonstrate beam steering metasurfaces with varied supercell periods to obtain different beam deflection angles. Second, we design and investigate a metalens with a focal length of 11.5 cm. Third, we exploit the orbital angular momentum of light and verify the mediated topological charges by revealing the spiral intensity pattern. Then, the phase profile of a metasurface hologram is designed with the Gerchberg-Saxton algorithm and combined with the phase profile of a magnifying lens, highlighting an easy way for combining multiple functionalities within a single metasurface. Cascading two different metasurfaces allows combining different functionalities by exploiting the unaffected incident light. Finally, we use a diffractive neural network for designing a single hologram metasurface featuring two different hologram patterns at certain distances behind the metasurface.

**Programming large-area geometric phase beam steering metasurfaces:**

The conventional fabrication of large area metasurfaces is complex and requires several lithography steps, incompatible with the requirements of rapid prototyping. Direct laser writing of metallic antennas with the plasmonic PCM IST instead offers a promising platform for rapid development of complex phase-modulated metasurfaces. The large-area metasurfaces investigated are directly optically written in 100 nm amorphous IST on top of a transparent $CaF_2$ substrate with precise laser pulses (see **Figure 1A**). In particular, we applied the direct laser writing system Photonic Professional GT from *Nanoscribe* equipped with highly precise galvo mirrors to redirect the laser beam on the sample and induce the crystallization process (see **Methods**). While amorphous IST exhibits dielectric behavior with a constant permittivity of 14, its crystalline phase follows a Drude-like behavior (ε<0) in the entire infrared spectral range.[24] The permittivity of amorphous and crystalline IST is shown in **Supplementary Note 1**. Consequently, it is possible to directly program entire large-area metasurfaces by locally crystallizing spatially varying nanoantennas within the amorphous IST. Full 2π-phase control within the metasurfaces fabricated is achieved by rotating the antennas from zero to 180 degrees and illuminating the metasurface with circularly polarized light. This concept is also known as Pancharatnam-Berry phase or geometric phase. The rotation angle $β$ directly translates to the phase of the scattered light $ϕ$ via:

$$ϕ = 2 \cdot β \quad (1)$$

The phase is only controlled by the rotation of the antenna and the chirality of the scattered light is reversed with respect to the incident light, allowing for clear distinction of the incident light from the scattered light.[34] A more detailed description of this geometric phase concept can be found in **Supplementary Note 2**.

The employed antennas featuring a length of 2.5 µm are designed to be resonant at a wavelength of 9 µm which is the operation wavelength of the infrared quantum cascade laser and the employed quarter-wave plates. Note, that this choice of wavelength does not display a limit of the demonstrated concept and any other infrared wavelength would be also possible. Measured transmittance spectra of crystallized IST antennas for different lengths can be found in **Supplementary Note 3**.

First, we investigate two beam steering metasurfaces with spatially varying antennas along the supercell period Γ. Within the supercell period, the phase gradient varying from 0 to 2π is determined by the rotation angle of the antennas. Engineering the supercell period Γ leads to the deflection angle via[3]:

$$\theta = \arcsin\frac{\lambda}{\Gamma} \quad (2)$$

The operation wavelength is set to 9 µm and two metasurfaces with supercell periods of 18 µm and 36 µm are designed, leading to theoretically calculated deflection angles of 30° and 14.5°, respectively. The period between individual antennas is set to 4 µm with an antenna length of 2.5 µm. Light microscope images of the optically crystallized metasurfaces are shown in **Figure 1B**. Note that for the beam steerer with a corresponding supercell period of 18 µm, two adjacent supercells are displayed. The entire metasurface with a size of 4x4 mm² consists of 1 million individual antennas with varied orientation fabricated within 2.5 hours. The phase difference introduced between adjacent antennas is given by 2π/N, with N referring to the number of antennas within the supercell.

Afterwards, we characterize the deflected beam transmitted through the metasurfaces with a home-build setup (see **Methods**). The incident LCP beam is transmitted through the metasurface, while the scattered RCP beam is deflected according to formula 2. The detector is rotated to measure the angle-resolved beam intensity. Applying a second quarter-wave plate combined with a linear polarizer allows for clear distinction between both polarization chiralities.

**Figure 1C** displays the measured laser intensities after passing both beam steering metasurfaces dependent on the angle and polarization chirality. The initial LCP chirality (dashed curves) is transmitted through the metasurface, showing maximum intensity at 0° deflection. In contrast, light with the opposite RCP chirality (solid lines) is deflected according to formula (2). Here, the metasurface with a supercell period of 18 µm exhibits a peak for the deflected RCP light at 30°, while the metasurface with the larger supercell period of 36 µm reveals a deflection angle of 14.5°. Remarkably, the experimentally measured beam intensities are in very good agreement with numerical far-field simulations (see **Supplementary Note 4**). The efficiency of the metasurfaces is determined by comparing the intensity of the deflected light with the unaffected transmitted light, revealing values around 10 %.

We performed field simulations of the RCP electric field transmitted through the metasurface for incident LCP light as shown in **Figure 1D**. The RCP component is deflected by 30° for the metasurface with a supercell period of 18 µm, and by 14.5° for the metasurface with a supercell period of 36 µm, respectively. Hence, the experimentally obtained deflection angles are well reproduced with electric field simulations and numerical far-field simulations.

Moreover, our designed beam steering metasurface is robust against fabrication imperfections due to the broad electric dipole resonances of the IST antennas. We demonstrate that even length variations of ±20 % perform similarly without significant decrease in performance (see **Supplementary Note 5**).

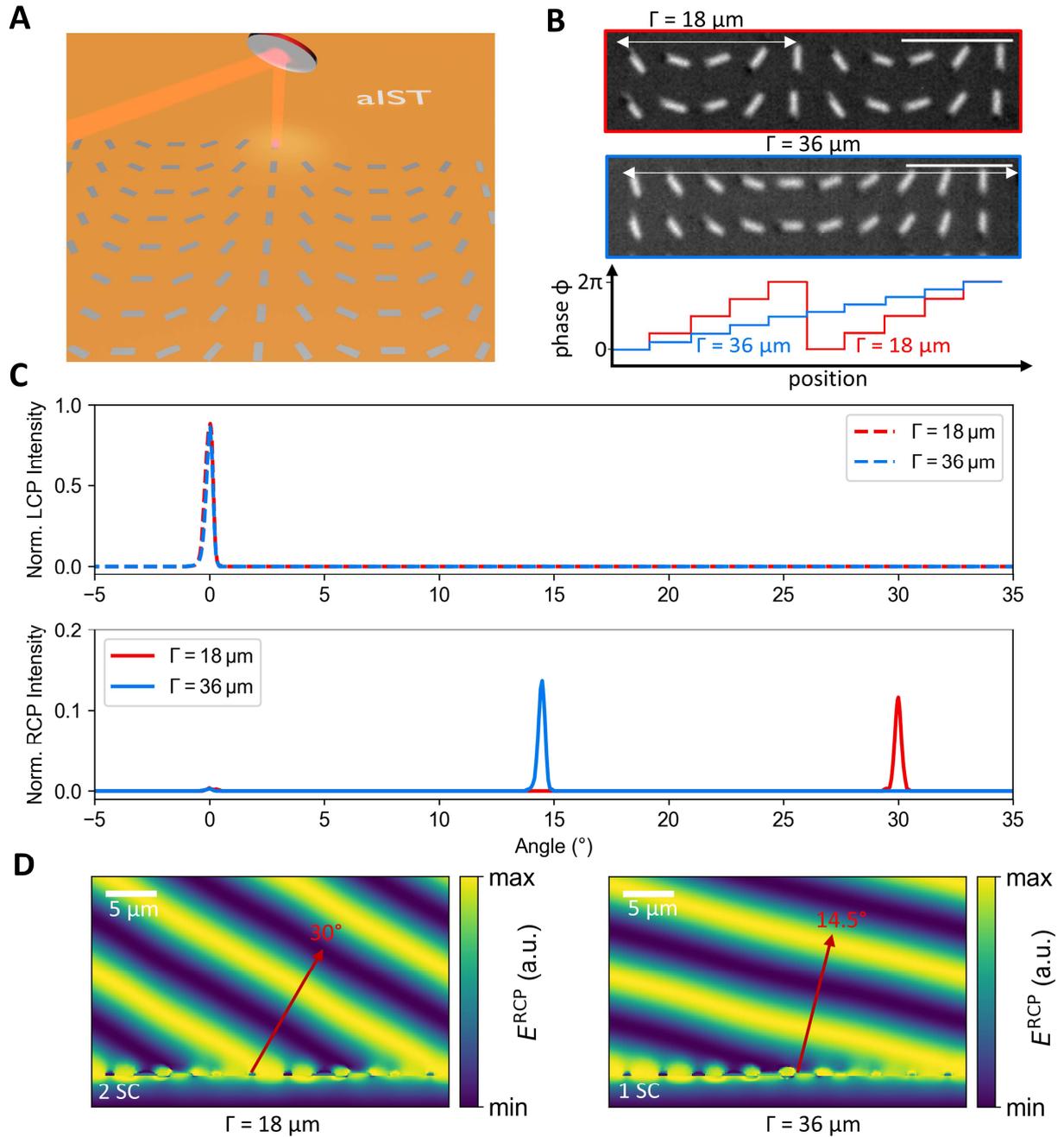

**Figure 1: Beam steering Metasurface with IST. A)** Concept of programming large-area geometric metasurfaces by optically crystallizing rotated IST rod antennas. Controlling the rotational angle of the rod antennas results in full 2π phase control of the metasurface. **B)** Light microscope images of two different beam steering metasurfaces with supercell periods Γ of 18 µm and 36 µm. The phase increases from 0 to 2π along the supercell period. The scale bars are 10 µm. **C)** Measured beam intensities for two different beam steering metasurfaces with supercell periods of 18 µm and 36 µm. The RCP component of the circularly polarized light is deflected, while the initial LCP light is not affected. **D)** Simulated RCP polarized electric field for incident LCP polarization for the two metasurfaces. The measured beam deflection is well reproduced for both supercell (SC) periods.

**Focusing infrared radiation with a metalens:**

In addition, we investigate a metalens consisting of rotated crystalline IST rod antennas which focuses the converted RCP radiation to a focal spot 11.5 cm behind the metasurface (see sketch in **Figure 2A**). The applied phase pattern of the metasurface is shown in **Figure 2B** featuring concentric rings of equal phases calculated with:

$$\phi(r) = \frac{2\pi}{\lambda}(\sqrt{r^2 + f^2} - f). \tag{3}$$

Here, $f$ denotes the focal length of 11.5 cm and the operation wavelength λ of the metasurface is again set to 9 µm, while $r$ determines the radial antenna position.

**Figure 2C** displays a photograph of the fabricated 8x8 mm² large-area metasurface with each antenna resembling a fixed phase value. The period between adjacent antennas is set to 4 µm, and the exact orientation of each antenna at different positions on the metasurface is calculated with equation 4, leading to a nearly continuously varying phase pattern. The entire metasurface consists of 4 million single antennas fabricated in 8 hours. The same procedure is used for all following metasurfaces.

The corresponding intensity distribution measured in the xz-cross-section is shown in **Figure 2D** (see Methods about the measurement procedure). A clear focal spot with maximum intensity is observed 11.5 cm from the metasurface. Our experimental results are supported with numerical simulations done with the Python package LightPipes by simulating the propagation of the incident gaussian beam transmitted through the metasurface with the phase pattern of Figure 2B. The simulations shown in **Figure 2E** validate the focal spot at 11.5 cm from the metasurface with a comparable beam waist diameter in the focal spot. Finally, we investigate the diameter of the focal spot in **Figure 2F** exhibiting a beam waist of 189 µm by fitting a gaussian function to the measured intensity. The large beam waist is caused by the intrinsic ultra-low numerical aperture (NA) of 0.03 due to the large focal length of the metasurface. The low NA is only chosen for simplified measurement of the focal spot. Our approach of patterning IST metasurfaces allows for metalenses with larger NA too. The retrieved beam parameter product of the laser after passing the metasurface is close to the optimal value, confirming that our metasurface preserves the intrinsic beam quality (see **Supplementary Note 6**).

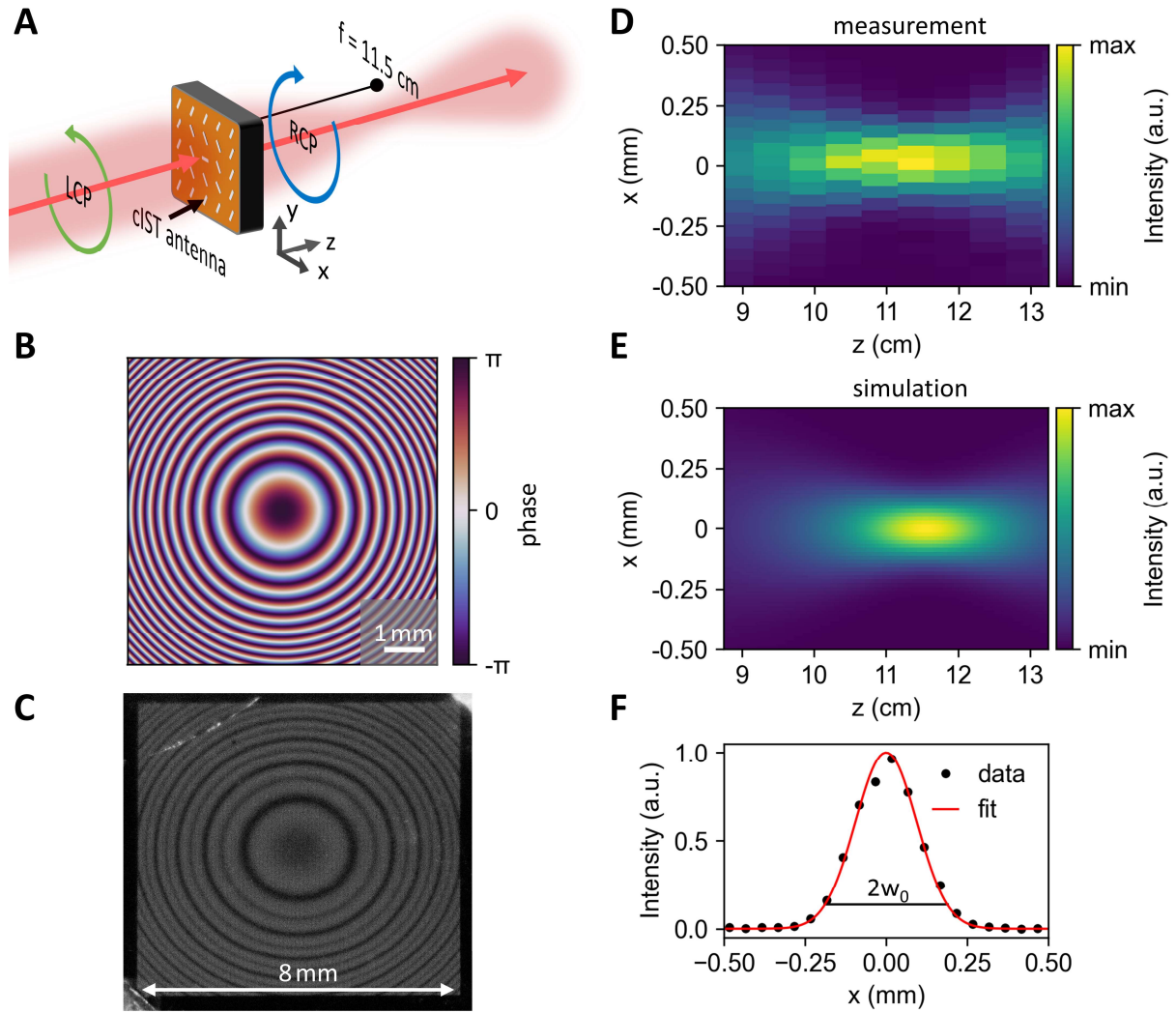

**Figure 2: Metalens with IST. A)** Schematic sketch of the working principle applied. The incident LCP light is converted to RCP radiation and focused according to the designed focal length. **B)** Imprinted phase profile of the metalens consisting of concentric rings with equal phases. **C)** Photograph of the fabricated 8x8 mm² metasurface consisting of 4 million rotated IST antennas. **D)** Measured intensity distribution in the xz-cross section clearly displaying a focal spot 11.5 cm from the metasurface. **E)** Simulated intensity distribution of a metasurface with the applied phase pattern from B. **F)** Cross-section of the measured focal spot revealing a beam waist $w_0$ of 189 μm.

**Beam-shaping exploiting orbital angular momenta:**

Generally, the concept of orbital angular momentum (OAM) has gained much interest in the past years.[35] The characteristic vortex beams consisting of different OAM modes feature a ring-like intensity distribution combined with helical phase factors $\exp(il\phi)$, with $l$ referring to the topological charge and $\phi$ the azimuthal angle. The orthogonal OAM modes can be superimposed to increase information capacity and boost optical communication systems.[36,37]

Here, we design three beam-shaping metasurfaces carrying an OAM in addition to the intrinsic spin angular momentum mediated by the polarization of the light. Therefore, we employ three helical phase patterns with topological charge *l* of one, three and five (see **Figure 3A**). As stated before, the orientation of the antennas is calculated according to the exact position onto the metasurface.

The corresponding far-field intensity measurements of our metasurfaces can be seen in **Figure 3B**. The diameter of the observed rings increases, pointing towards different topological charges. The measured intensity cross-section along the rings can be found in **Supplementary Note 7**. The determination of the intrinsic OAM is achieved by direct interference of the unaltered beam with the light carrying the OAM after passing the metasurface (see Methods for more details). The results are shown in **Figure 3C**. Here, the number of observable spiral arms is directly associated with the topological charge. The first image for l = 1 features only one spiral arm, while the second image for l = 3 features three spiral arms and so on, verifying the OAM carried by the photons after passing the metasurface.

Numerical simulations of a gaussian beam imprinted with the corresponding phase profiles in Figure 3A are shown in **Figure 3D**. Characteristic ring-like patterns associated with doughnut modes with increasing diameter for increasing topological charges appear with diameters comparable to the experimentally obtained images. **Figure 3E** displays far-field intensity simulations of the OAM beam superimposed by the incident gaussian beam, revealing the spiral arms as described before. The simulations are in very good agreement with the experimental data.

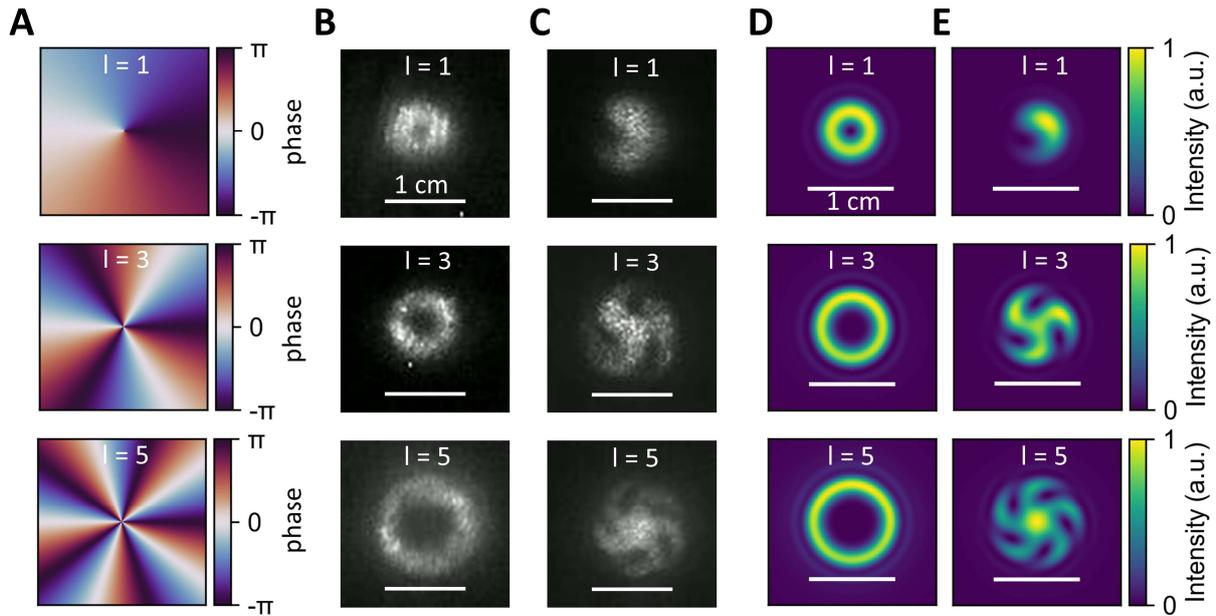

**Figure 3: Encoded orbital angular momentum. A)** Applied helical phase patterns carrying an orbital angular momentum of 1, 3 and 5 for three different metasurfaces. **B)** Measured far-field intensity patterns of the beam 60 cm from the metasurface displaying a ring-like structure with increasing diameter for higher orbital angular momenta. **C)** Direct interference of the unaffected incident light with the scattered light from the metasurface lead to a spiral pattern with the number of spiral arms revealing the orbital angular momentum. **D)** Simulated far-field intensity pattern of the helical phase profiles demonstrating similar ring patterns compared to B. **E)** Simulated spiral patterns due to the interference of the incident gaussian beam and the transformed beam after passing the metasurface reveal very good agreement between experiment and simulation.

**Enlarged infrared holography with a multifunctional metasurface:**

The vast flexibility of direct programming complex phase patterns with IST is demonstrated by designing a holographic phase pattern with the Gerchberg-Saxton algorithm (see Methods) to create a specific infrared hologram in the far-field. Because the size of the designed holographic image is small and therefore challenging to measure, we superimpose the retrieved phase pattern of the hologram with the phase pattern of a magnifying lens. This highlights the ability of combining multiple functionalities within a single metasurface and leads to simplified imaging of the resulting hologram with a size of several centimeters. The concept is sketched in **Figure 4A**. The targeted far-field intensity pattern is displayed in **Figure 4B** showing the letters 'ir nano' with uniform intensity. The final calculated phase pattern combined with the phase pattern of the lens for the designed hologram 60 cm from the metasurface can be seen in **Figure 4C**. Here, the phase values are simply added to each other. The photograph of the fabricated 8x8 mm² metasurface shows a similar pattern caused by scattering of the visible light (c.f. **Figure 4D**). The measured hologram in combination with numerical simulation is shown in **Figure 4E** for different distances behind the metasurface. While for small distances (e.g. 20 cm behind the metasurface) no clear image is observed, at larger distances around 60 cm behind the metasurface, the letters 'ir nano' are clearly resolvable. Due to the magnifying lens, the size of the hologram increases with further distance to the metasurface. The experimental data and the simulations agree well with each other.

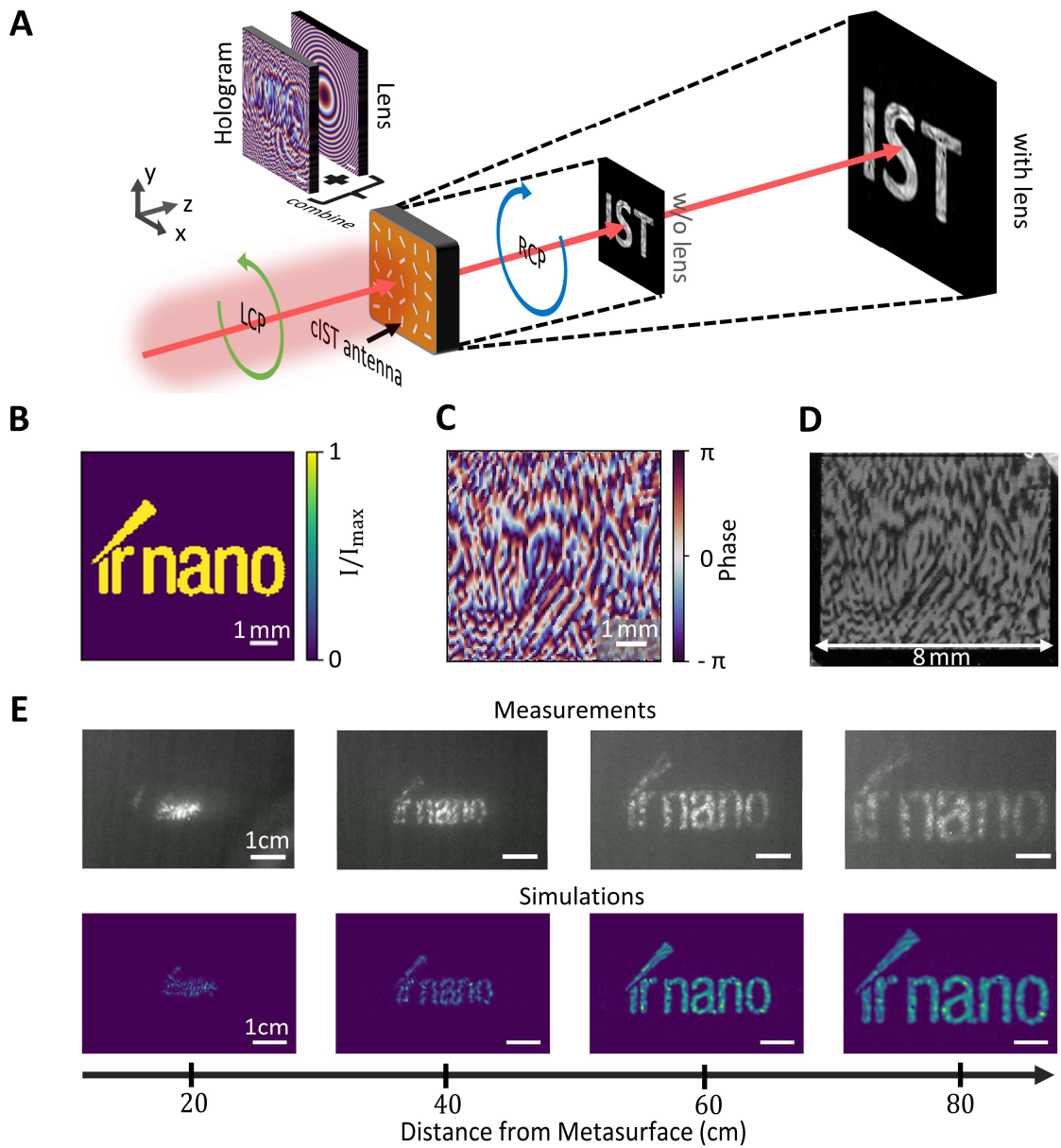

**Figure 4: IST Holography Metasurface. A)** Schematic sketch of combining the phase pattern of the hologram with the pattern of a magnifying lens to achieve enlarged imaging of the RCP polarized holographic image. **B)** Targeted far-field intensity pattern displaying the letters 'ir nano' at 60 cm

distance from the metasurface. **C)** Calculated metasurface phase profile with the Gerchberg-Saxton algorithm according to the targeted intensity pattern superimposed with the phase profile of a magnifying lens. **D)** Photograph of the 8x8 mm² metasurface written optically. **E)** Measured and simulated intensity patterns at varied distances from the metasurface. The designed hologram is best visible 60 cm behind the metasurface.

In another experiment, a third functionality of a beam steerer is added to the magnified holography metasurface by superimposing the previous phase mask with the phase mask of a beam steerer (see **Figure 5A**). The resulting metasurface deflects the enlarged hologram by a given angle of 10° as shown in **Figure 5B**. For better visualization, the incident LCP light is not filtered out completely, leading to a point-like intensity pattern at 0°. This demonstrates the ability of combining several different functionalities within the same metasurface by simply superimposing the corresponding phase masks which is not possible with conventional optical elements.

Moreover, by cascading two metasurfaces the remaining unaffected incident polarized light can be reused for another functionality. This is possible because the converted RCP light from the first metasurface transmits unaffected through the second metasurface. A sketch of the setup and concept employed is shown in **Figure 5C** by inserting another metasurface (black dotted line) into the beam path. The first metasurface employed corresponds to the OAM metasurface with l = 3, while the second metasurface features the deflected and enlarged hologram from Figure 5A. The resulting intensity profiles at the screen are displayed in **Figure 5D**. Here, the characteristic ring-pattern of the OAM appears at 0° and the deflected hologram at 10°. A second example of cascaded metasurfaces with two holograms is shown in **Supplementary Note 8**.

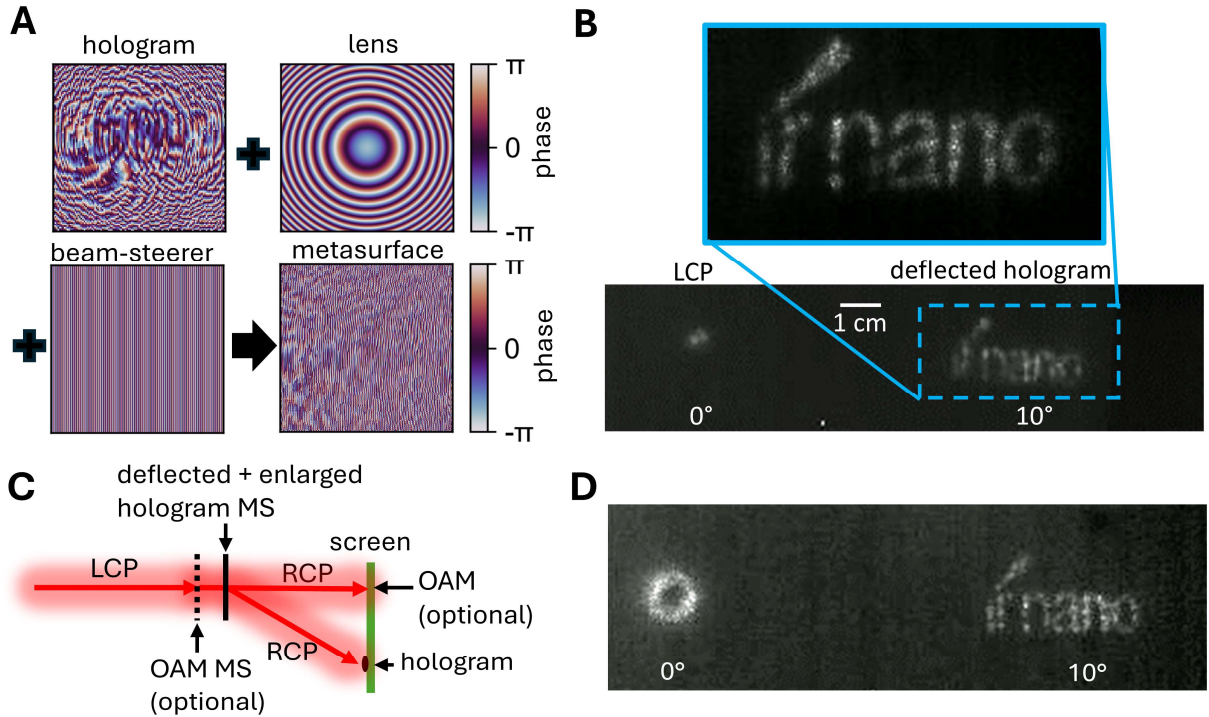

**Figure 5: Multifunctionality and cascaded metasurfaces. A)** Superimposed phase masks of the hologram, magnifying lens, and beam steerer leading to the phase mask of the metasurface for enlarged and deflected holography. **B)** Measured intensity profiles of the corresponding metasurface. The incident LCP light is attenuated and visible at 0°, while the 'ir nano' hologram appears deflected at 10°. **C)** Sketch of the measurement setup for cascaded metasurfaces creating the OAM at 0° (optional) and the deflected and enlarged hologram at 10°. **D)** Measured OAM intensity profile at 0° and deflected hologram at 10°.

Finally, we design a hologram metasurface exhibiting two different holograms at different positions $z_1$ and $z_2$ from the metasurface with a diffractive neural network (see **Supplementary Note 9**).[38,39] The phase of the scattered light is altered to not only display a hologram at a set distance, but also additionally enables reordering of the light upon propagation to form a second hologram with a different intensity distribution. This concept is visualized in **Figure 6A**. Here, at position $z_1$, the hologram showing a distorted lattice representing the amorphous phase with the caption 'aIST' is displayed. At a second position $z_2$, the observable hologram changes to a periodic lattice representing the crystalline phase with the caption 'cIST'. The calculated phase mask is shown in **Figure 6B**. Notice that we took the actual beam shape of the laser into account to achieve homogeneous intensity distributions within the hologram images. Measurements performed with a Pyrocam IV by *Ophir Photonics* at 16 cm and 21 cm are shown in **Figure 6C**. Remarkably, the observable pattern changes as designed upon increasing the distance. Numerical simulations (see **Figure 6D**) are in good agreement with the experimental data. The evolution of the hologram by varying the distance behind the metasurface is shown in **Supplementary Video 1**.

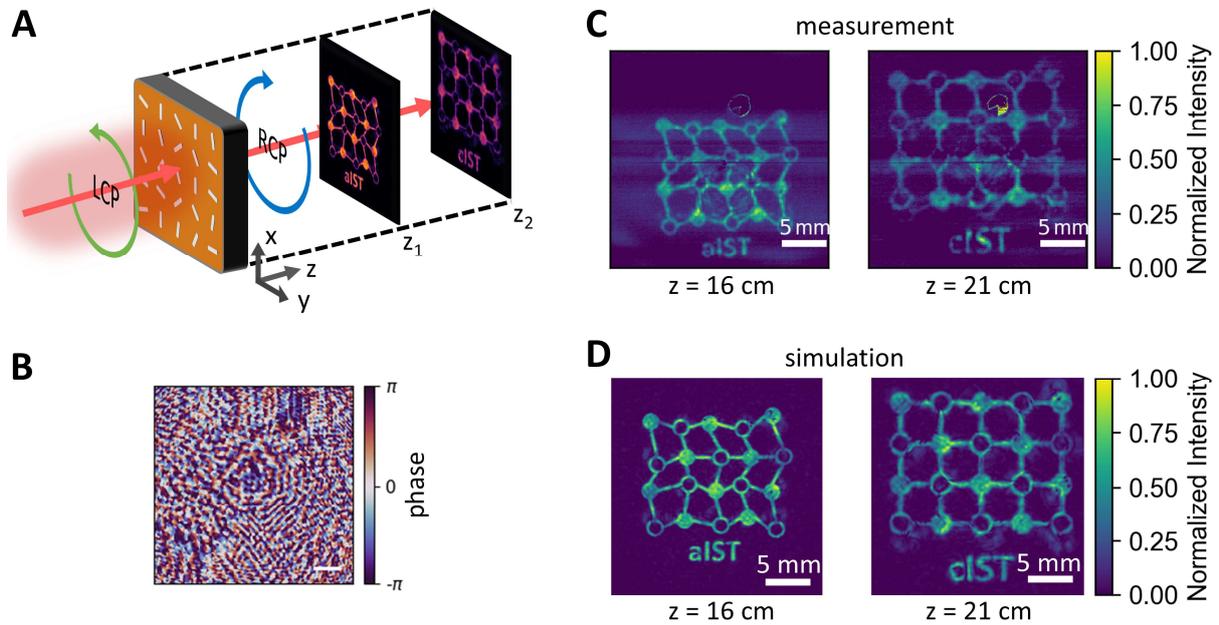

**Figure 6: Dual-hologram. A)** Schematic sketch of the employed dual-holography concept. The IST metasurface leads to two different holograms at specific distances behind the metasurface. **B)** Applied phase mask computed with a diffractive neural network. **C)** Measurement of the holograms displaying 'aIST' at $z_1$ = 16 cm with a distorted lattice structure and 'cIST' with a periodic lattice at $z_2$ = 21 cm. **D)** Numerical simulations clearly resolve the two holograms at different distances behind the metasurface.

**Conclusions:**

In summary, we demonstrated direct programming of geometric phase metasurfaces consisting of rotated crystalline IST rod antennas within the plasmonic PCM IST for infrared beam-shaping. Tailoring the phase of the metasurface gives access to numerous applications such as beam steering, lensing, and holography. The concept of directly programming metasurfaces with IST antennas is much simpler, cost-effective, and faster compared to cumbersome fabrication techniques such as conventional lithography involving multiple complex mask design and costly etching steps. The operation wavelength of the metasurfaces is only given by the length of the rotated antennas and can be easily scaled to target the entire infrared range. Commonly, the infrared spectral range displays a significant lack of commercially available beam shaping elements. Hence, our demonstrated concept paves the way towards rapid prototyping and simple production of reconfigurable meta-optics in the infrared even for industrial purposes.

Customized infrared metasurfaces can be employed in security applications, beam shaping for material processing, or quantum manipulation and information encoding via exploiting the orbital angular momenta of photons.[40,41] Reprogramming the metasurface and modifying the phase profile features a convenient way for replacing spatial light modulators and diffractive optical elements.[42]

**Methods:**

**Sample fabrication:**

Direct current magnetron sputtering is used to deposit a 100 nm thin amorphous $In_3SbTe_2$ film on top of 1x1 cm² infrared transparent $CaF_2$ substrates. Afterwards, a 50 nm thin layer of $(ZnS)_{80}:(SiO_2)_{20}$ is deposited with radio frequency magnetron sputtering to prevent the samples from oxidization and to facilitate the crystallization process as an anti-reflection coating for the switching laser.

**Optical switching:**

Local crystallization of the individual crystalline IST antennas is done with the direct laser writing system Photonic Professional GT from *Nanoscribe*. Here, a 100 fs pulsed laser with a central wavelength of 780 nm and a repetition rate of 80 MHz is employed. The laser pulses are focused by a 63x objective with a numerical aperture of 1.4 onto the sample. The high numerical aperture is achieved by employing the oil 3-(Trimethoxysilyl)propyl methacrylate. Precise movements of the laser are enabled by galvo mirrors with a writing field of 100 x 100 µm². Coarse movements of the sample are retrieved by a motorized stage (range of several cm).

For all metasurfaces, we operated the system in the Continuous Mode with a scan speed of 3500 µm/s and a laser power of 15 mW.

The writing time for the beam steering metasurfaces consisting of 1 million antennas was 2.5 hours, while the writing time of the metasurfaces with 4 million antennas took about 8 hours.

**FTIR measurements:**

The measured transmittance spectra are recorded by a Bruker Vertex 70 interferometer connected to a Bruker Hyperion 2000 microscope. We applied a 15x Cassegrain objective with a numerical aperture of 0.4 featuring an angular distribution from 10 to 24 degrees. The spectra are recorded with 1000 scans and a resolution of 4 $cm^{-1}$.

**Metasurface characterization:**

The fabricated metasurfaces are characterized by a home-built setup. A quantum cascade laser from Daylight Solutions with a wavelength of 9 µm and vertical polarization is circularly polarized with a quarter- wave plate from Optogama designed for a wavelength of 9 µm and then directed onto the metasurface. The initial chirality is filtered out by a second quarter-wave plate in combination with a linear polarizer after passing the metasurface. For detection, different systems were employed. The beam steering metasurfaces (see Figure 1) are characterized with a mercury cadmium telluride (MCT) detector positioned on a mechanical micrometer stage movable along a semi-circle. For the measurements of the metalens (see Figure 2), we employed a knife edge razor blade positioned on a micrometer stage at different positions behind the metasurface and measured the change in the detected laser power. The simple hologram and the orbital angular momentum metasurfaces (see Figure 3 and 4) are projected on a screen and imaged with a FLIR T335 thermal camera. The spiral intensity pattern of the OAM metasurface caused by the direct interference with the incident light is obtained by rotating the linear polarizer to achieve approximately similar laser powers of the converted RCP and incident LCP light. For the dual-hologram (see Figure 6), we employed the Pyrocam IV by Ophir Photonics to measure the beam intensity profile after passing the metasurface directly. Schematic sketches and more detailed explanations of the different measurement setups can be found in **Supplementary Note 10**.

**Simulations:**

Numerical simulations of the antenna spectra and the beam steering metasurface are done with the commercially available program CST Studio Suite from Dassault Systems. The permittivity of IST is taken from Supplementary Note 1. For the $CaF_2$ substrate and the capping layer, a constant refractive index of 2.1 and 1.4 are assumed, respectively. Floquet Mode Ports are chosen to excite the simulated structures. Unit cell boundaries in lateral dimensions and open boundaries in vertical dimensions are applied. The simulations for the metalens, the hologram and the orbital angular momentum metasurface are done with the freely available Python package LightPipes. Here, the phase profile of the respective metasurface is impinged by a gaussian beam and then the far-field intensity is calculated for varied distances behind the metasurface. The divergence of the incident laser beam in the experiment is taken into account. The calculation of the dual-hologram metasurface is described in **Supplementary Note 9**.

**Acknowledgements**

L.C., A.M., R.S. and T.T. conceived the research idea; L.C. and F.B. designed the research; F.B. and A.M. carried out the optical switching; F.B. performed the metasurface measurements. L.C. and F.B. analyzed the data and carried out the numerical simulations. P.B. and C.H. calculated the phase mask of the Dual-hologram. M.W. provided the sputtering equipment and phase-change material expertise; all authors contributed to writing the manuscript. The authors thank Maike Kreutz for the sputter deposition of the thin film layer stack.
The authors acknowledge support by the Deutsche Forschungsgemeinschaft (DFG No. 518913417 & SFB 917 "Nanoswitches").

**Data availability**

All data are available in the article and the Supplementary Information, and are available from the corresponding authors on reasonable request.

**Code availability**

The source code for the calculations conducted in this study is available from the corresponding authors on reasonable request.

# Supporting Information for

**Infrared Beam-shaping on Demand via Tailored Geometric Phase Metasurfaces employing the Plasmonic Phase-Change Material In$_3$SbTe$_2$**

*Author(s), and Corresponding Author(s)\**

*Lukas Conrads[+,*,1], Florian Bontke[+,1], Andreas Mathwieser[2], Paul Buske[3], Matthias Wuttig[1], Robert Schmitt[2], Carlo Holly[3,4], Thomas Taubner[#,1]*

+ both authors contributed equally

**Affiliations**

[1] Institute of Physics (IA), RWTH Aachen University, D-52056 Aachen, Germany

[2] Fraunhofer Institute for Production Technology IPT, 52056 Aachen, Germany

[3] Chair for Technology of Optical Systems, RWTH Aachen University, 52056 Aachen, Germany

[4] Fraunhofer Institute for Laser Technology ILT, 52056 Aachen, Germany

\* Email: conrads@physik.rwth-aachen.de

# Email: taubner@physik.rwth-aachen.de

**This PDF file includes:**

**Supplementary Note 1: Dielectric Function In$_3$SbTe$_2$**

The plasmonic phase-change material (PCM) In$_3$SbTe$_2$ (IST) features a dielectric amorphous phase and a metallic crystalline phase. The permittivity of IST for the mid-infrared spectral range is shown in **Figure S1**. The permittivity of crystalline IST follows a Drude-like behavior with a negative real part of the permittivity (ε' < 0). Details about modelling the permittivity can be found elsewhere.[1]

Hence, by locally crystallizing IST with precise laser pulses, large-area metasurfaces can be directly programmed within a dielectric surrounding.

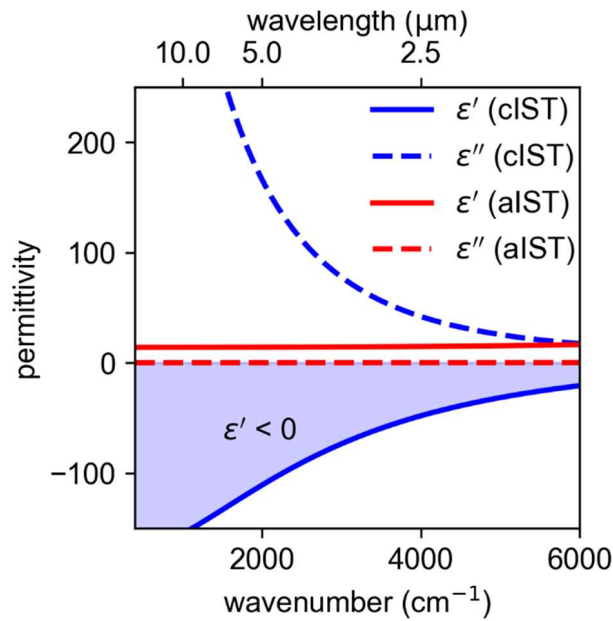

**Figure S1**: Dielectric function of amorphous (red) and crystalline (blue) IST. The permittivity of crystalline IST follows a Drude-like behavior, while the permittivity of amorphous IST is constant with a value of 14.

## Supplementary Note 2: Geometric Phase Metasurfaces

Exploiting rotated metallic antennas in combination with circularly polarized light provides a convenient way in tailoring the phase of the employed metasurfaces at will.

In particular, the transmitted light for incident left-handed circularly polarized (LCP) light $E_T^L$ through such a metasurface consisting of anisotropic scatterers, e.g. rod antennas, is given by:[2]

$$E_T^L = \frac{t_o+t_e}{2} E_I^L + \frac{t_o-t_e}{2} \exp(im2\beta) E_I^R \quad (1)$$

Here, the $t_o$ and $t_e$ refer to the scattering coefficients for incident linearly polarized light along the two principal axes, and m describes the handedness of the circular polarization, being ±1. Hence, the transmitted light features a component with the same handedness of the incident polarization $E_I^L$ and a component with opposite handedness, i.e. right-handed circularly polarized (RCP) light $E_I^R$ with an additional phase of m2β, also called Pancharatnam-Berry Phase. The phase is only controlled by the rotation of the antenna and the chirality of the scattered light is reversed with respect to the incident light, allowing for clear distinction of the incident light from the scattered one. In particular, the phase $\phi$ of the scattered light depends on the rotation angle $\beta$ via:

$$\phi = 2 \cdot \beta \quad (2)$$

A visualization of this concept is shown in Figure S2.

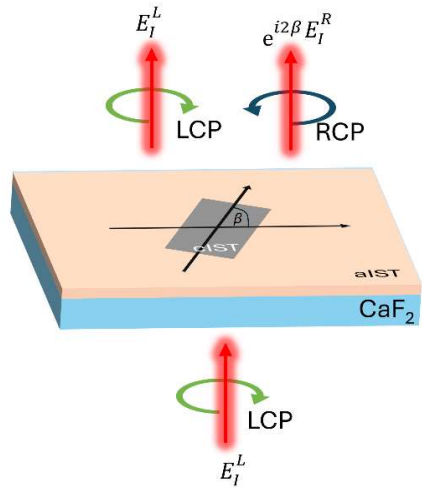

**Figure S2: Concept of geometric phase metasurface.** The transmitted light features a component with the same handedness of the incident polarization and a component with opposite chirality and an additional phase dependent on the rotation angle of the antenna.

## Supplementary Note 3: Transmittance spectra IST antennas

The measured transmittance spectra of crystalline IST rod antennas with varied antenna length with light polarized parallel to the long antenna axis are shown in **Figure S3**. A broad minimum in the transmittance corresponding to the electric dipole resonance of the rod antennas occurs. The broad antenna resonances are beneficial for broadband operation, not limiting the operation wavelength of the metasurface which conventionally is the case for narrow dielectric resonators. For increased antenna length, the electric dipole resonance shifts towards larger wavelengths. Antenna arrays with a length of 2.5 µm display an electric dipole resonance at 9.1 µm and are therefore used for the different metasurfaces. The results are in good agreement with previous experiments about the resonance position of crystalline IST antennas.[1]

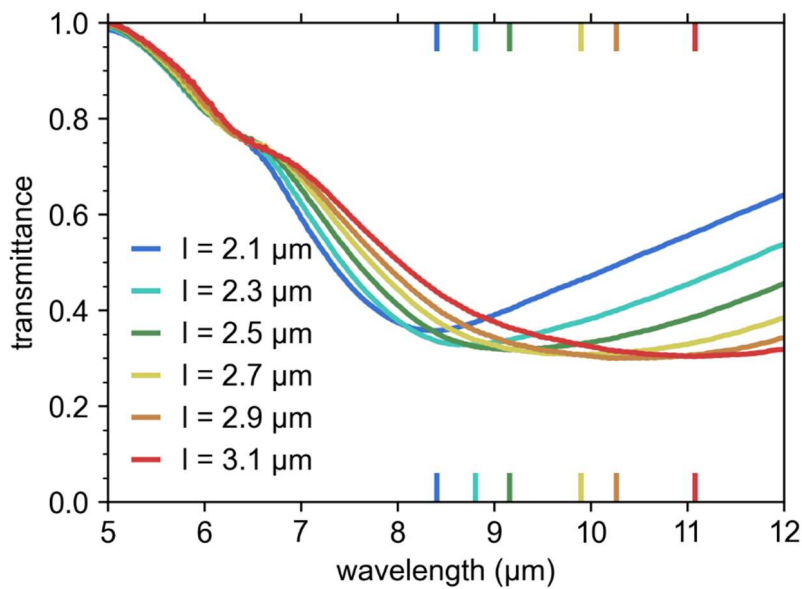

**Figure S3: Measured FTIR transmittance spectra of crystalline IST antennas with varied antenna lengths.**

**Supplementary Note 4: Measured and simulated beam steering metasurfaces**

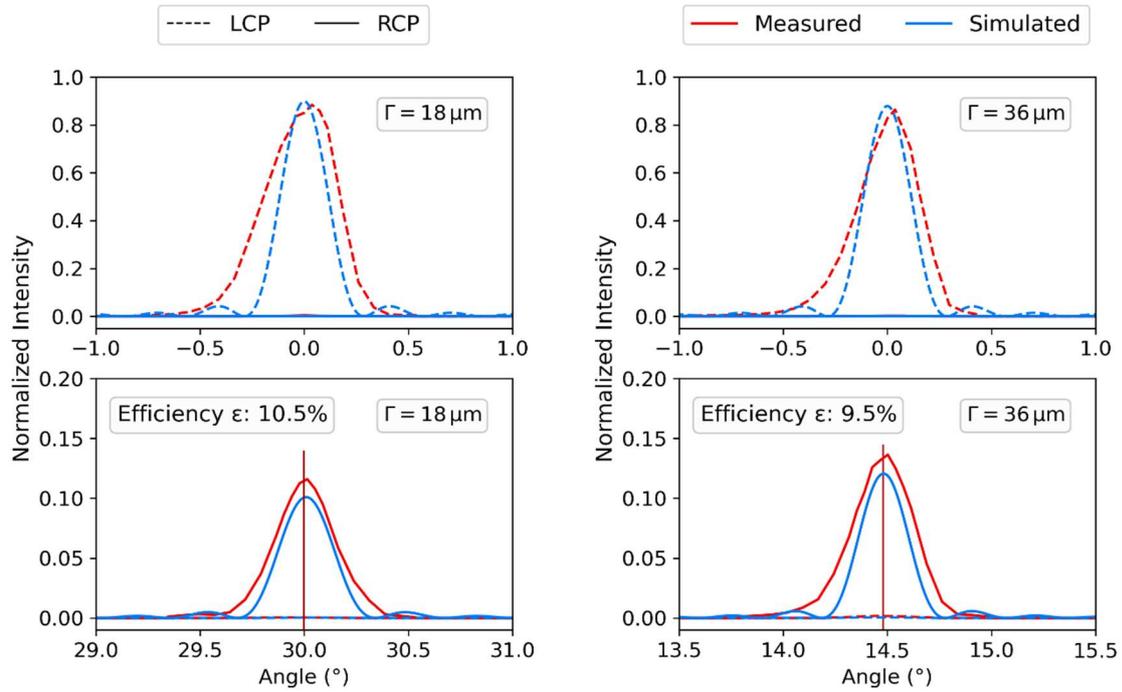

**Figure S4: Measured and simulated beam intensities after passing the two beam steering metasurfaces.**

## Supplementary Note 5: Metasurface Robustness

Commonly, metasurface fabrication requires high precision and accuracy of the fabricated nanoantennas without significantly decreasing the metasurface performance. Since the electric dipole resonances of the employed IST antennas are very broad at 9 µm with a full-width-half-maximum of 4.4 µm, we demonstrate robustness against fabrication imperfections indicating broad band performance of our metasurfaces. Therefore, we fabricate three different beam steering metasurfaces with a supercell period of 18 µm and modified antenna lengths varied from 2.2 µm to 2.9 µm. Light microscope images of the supercells for the different metasurfaces are shown in **Figure S5A**. The corresponding measured transmittance spectra of antenna arrays with varied lengths (see **Figure S5B**) display broad electric dipole resonances at 8 µm for antennas with a length of 2.2 µm shifted to 10.5 µm for antennas with a length of 2.9 µm. Since the applied quarter-wave plates limit the operation wavelength of the metasurfaces to 9 µm, we compare the beam deflection efficiencies for varied antenna lengths as displayed in **Figure S5C**. Because we keep the supercell period constant, the deflection angle is fixed to 30°. The general performance of the metasurfaces feature normalized beam deflection efficiencies around 10 %, despite of the strongly differing resonance positions. Moreover, the deflection intensities of the beam steerer with largest antenna lengths exceeds the optimum case with respect to the resonance position at the operation wavelength. This behavior is further investigated by numerical far-field simulations of the scattered electric field in the xz-cross-section for single antennas (c.f. **Figure S5D**). Here, the lower hemisphere corresponds to scattered light in the substrate closely related to the transmitted light. Remarkably, the antennas with a length of 2.9 µm exhibit more pronounced scattering in the forward direction (lower hemisphere) instead of the backward direction (upper hemisphere) compared to smaller antennas. This explains the difference in the measured beam deflection intensities for varied antenna lengths.

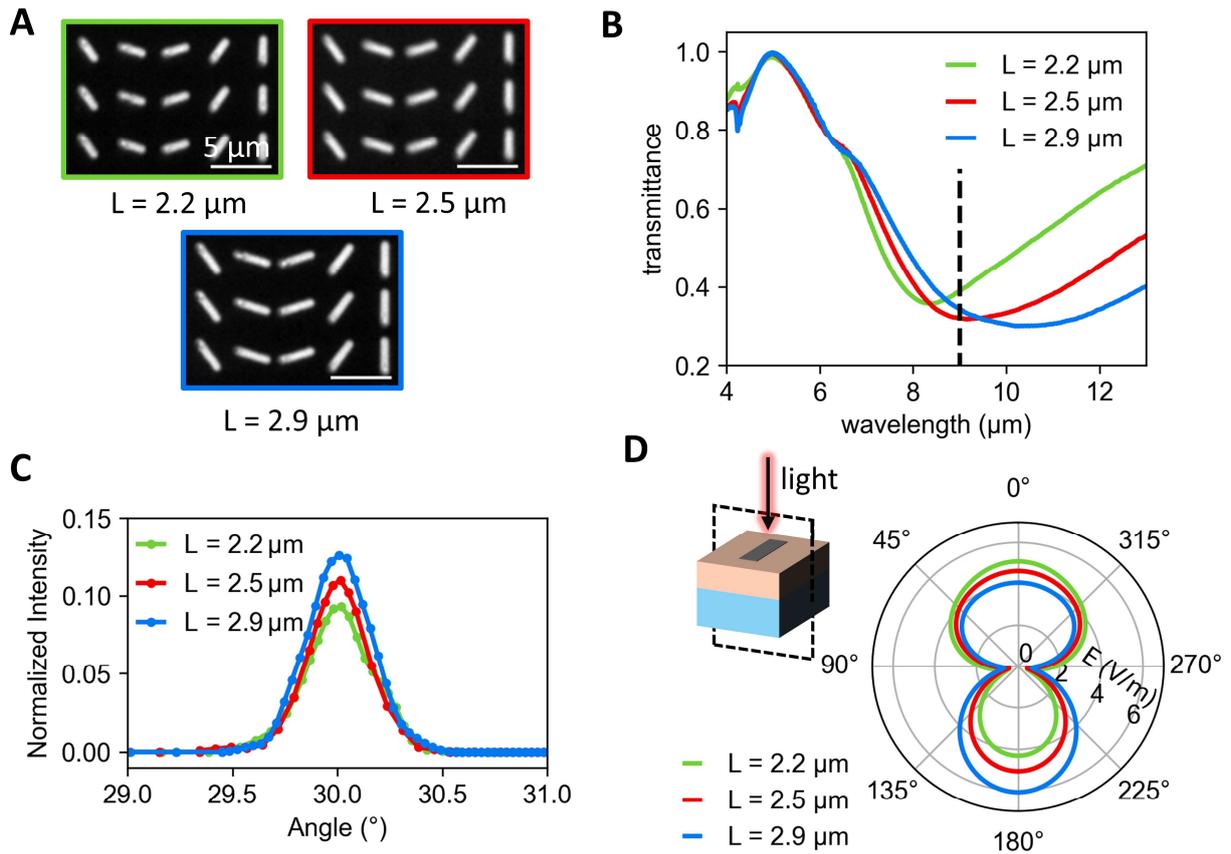

**Figure S5: Beam steering performance for varied antenna length. A)** Light microscope images of three different beam steering metasurfaces with varied antenna lengths from 2.2 µm to 2.9 µm. **B)** Measured transmittance spectra of antenna arrays with the three different antenna lengths with electric dipole resonances from 8 to 10.5 µm. The black dashed line indicates the operation wavelength of 9 µm. **C)** Measured beam deflection intensities for the different metasurfaces consisting of 3 different antenna lengths. **D)** Simulated far-field electric field scattering in the xz-cross-section (see inset) showing enhanced field scattering for the 2.9 µm long antenna.

## Supplementary Note 6: Beam quality after passing the metalens

The measured beam waist of the infrared laser after passing the metalens is shown in **Figure S6**. We fitted the evolving beam width with the formula $w(z) = w_0\sqrt{1 + \left(\frac{z}{z_R}\right)^2}$ to determine the beam quality $M^2 = \pi \frac{w_0^2}{z_R}$ after the metasurface. Here, $z_R$ refers to the Rayleigh length and $w_0$ refers to the beam waist in the focal spot. From the fit, we obtain a beam quality M² of 0.97 ± 0.03, which is very close to the optimal case of M² = 1. Therefore, we conclude that our metasurface even improves the intrinsic beam properties (see **Figure S6.2**). The unphysical value of M² smaller than one might be caused by imprecision during the measurements.

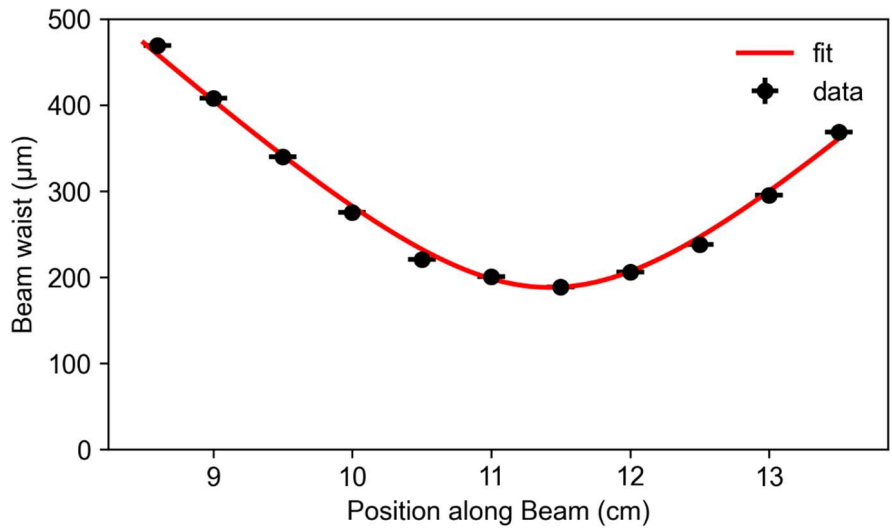

**Figure S6.1: Beam waist after passing the metalens.**

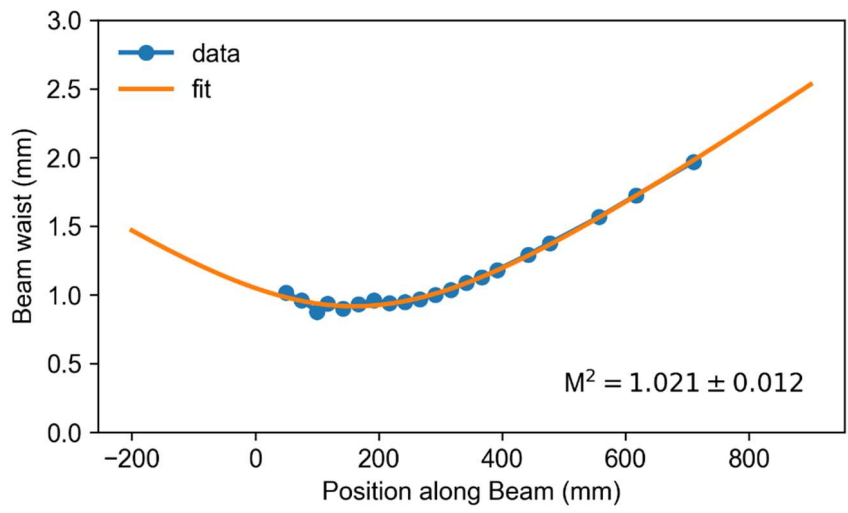

**Figure S6.2: Initial beam characteristic exhibiting an M² value of 1.021.**

**Supplementary Note 7: Intensity cross-section orbital angular momentum**

We measured the intensity of the shaped laser beam after passing the orbital angular momentum (OAM) metasurfaces with the knife-edge method to determine the diameter of the observed ring-like intensity patterns (see **Figure S7**). For the metasurface with encoded OAM of l = 1, the ring exhibits a diameter of 1.8 mm, while the ring obtained with the metasurface of encoded OAM of l = 3 features a diameter of 2.6 mm and for l = 5, the ring has a diameter of 3.4 mm.

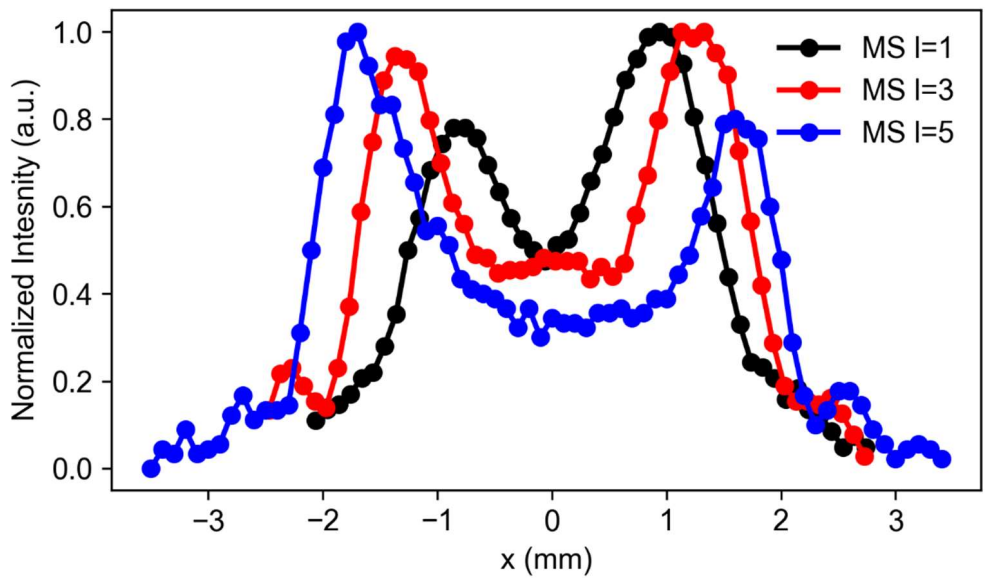

**Figure S7: Intensity cross-section through OAM metasurface at a distance of 15 cm.**

# Supplementary Note 8: Cascaded Metasurfaces

As already mentioned previously, the conversion efficiency of our applied metasurfaces is in the range of 10% (c.f. Figure 2 in the main manuscript and Supplementary Note 2). Hence, it is possible to cascade different metasurfaces and take advantage of the non-affected incident polarization of the light after passing through the first metasurface. This concept is visualized in **Figure S8A**. The first metasurface leads to a hologram as already demonstrated in Figure 4 in the main text. Afterwards, the incident LCP polarized light passes the first metasurface and hits the second metasurface designed for a second hologram which will be deflected due to the superimposed phase profile of a beam steerer. The incident LCP light is filtered out. The resulting measurement on the screen is shown in **Figure S8B**. At 0°, the first hologram is visible. Adjacent at 10° deflection angle, the second 'ir nano' hologram appears. Note that the order of the metasurfaces can be interchanged without affecting the result.

Therefore, the unaffected incident polarized light can be further used by multiple metasurfaces for different functionalities.

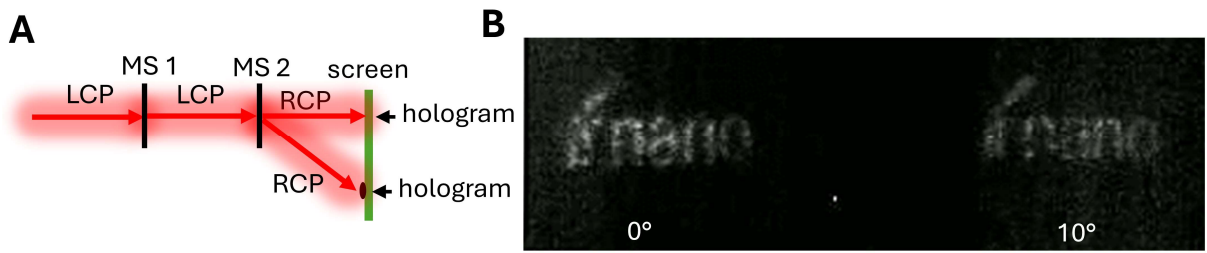

**Figure S8: Cascaded metasurfaces. A)** Schematic sketch of measuring two cascaded metasurfaces. The first metasurface (MS1) is designed to create a hologram. The non-affected incident LCP light is further directed to the second metasurface (MS2) designed for a second hologram deflected by 10°. **B)** Measurement of the screen showing the first hologram at 0° and the deflected hologram at 10°.

## Supplementary Note 9: Diffractive Neural Networks for Dual-hologram design

We design the phase mask of the dual-hologram by treating it as a single layer diffractive neural network[3], which is a stochastic gradient descent approach. This approach utilizes the analytical gradients provided in machine learning computation infrastructures by treating phase masks as trainable layers. This means that the entire optical system can be conceptualized as a physical neural network, with the input beam's complex amplitude as the input layer, the metasurface as the (here single) trainable layer and two output layers. The phase value of each pixel acts as an equivalent to a neuron in conventional neural networks. The connections between the layers are realized through Huygens' principle, so each "neuron" is connected to all previous and following nodes through spherical wave propagation. After implementing all the optics, the training can be performed exactly as a conventional neural network in an arbitrary neural network training environment - in this case in Pytorch Lightning. More details about the method are provided in ref[3] and ref[4]. We choose two separate target planes with two different target intensity distributions at distances of 16 cm and 21 cm respectively, measuring from the position of the metasurface plane. As the target distributions are significantly larger than the metasurface, the second distribution is scaled in its spatial extension to approximately fit to the divergence resulting from expanding the incoming beam to the size of the first target distribution.

Additionally, we employ two regularization techniques[5]. First, weight decay is used to penalize excessively large phase weights. Second, Laplacian regularization minimizes the second derivative between adjacent pixels. This encourages continuous and smooth phase masks, facilitating manufacturing and reducing speckle noise in the calculated intensity distributions.

The Laplacian regularization is implemented as a convolution with a Laplacian kernel

$$K = \begin{matrix} 0 & 1 & 0 \\ 1 & -4 & 1 \\ 0 & 1 & 0 \end{matrix}$$

so that the second derivative of the phase mask $\phi_{\text{mask}}$ can be expressed as

$$\Delta \phi_{\text{mask}} = K \cdot \phi_{\text{mask}}$$

with the Laplace operator $\Delta$. Then, the total loss function is given as

$$L_{total} = \sum_{n=0}^{1} \|I_n - I_n^{Target}\|^2 + \lambda_1 \|\phi_{mask}\|^2 + \lambda_2 \|\Delta\phi_{mask}\|^2$$

with $I_{0,1}$ as the respective two intensity distribution predictions and $\lambda_i$ as regularization hyperparameters.

The target intensity distributions displaying 'aIST' and 'cIST' with different lattice structures at the different image planes are shown in **Figure S9A**. The simulated far-field intensity distributions of the obtained metasurface phase mask are shown in **Figure S9B**, revealing good agreement with the target designs.

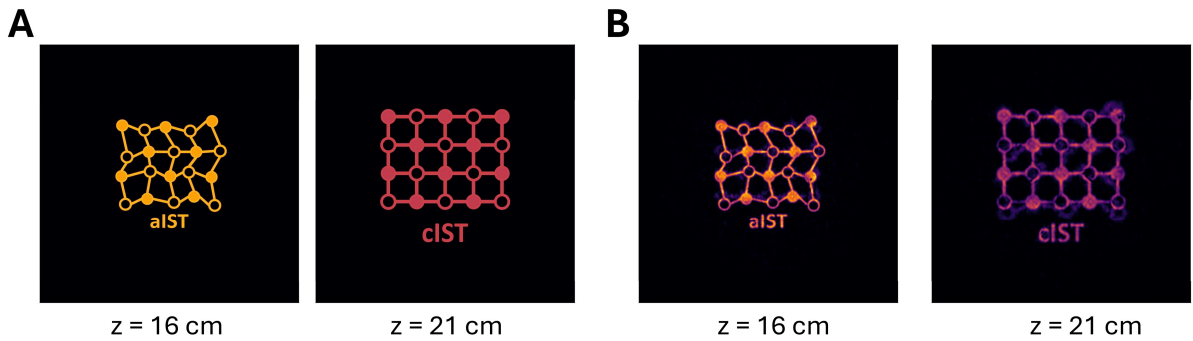

**Figure S9: Targeted (A) and simulated far-field intensity distribution of the dual-hologram metasurface.**

**Supplementary Note 10: Measurement Setups**

For characterization of the different fabricated metasurfaces, we employed different measurement setups (see **Figure S10**). For all setups, a quantum cascade laser from Daylight Solution with an operation wavelength of 9 µm is attenuated with a variable step attenuator (VA) and then directed to a quarter wave plate (QWP) to create left-handed circularly polarized (LCP) light. After transmitting the metasurface, the light features two components: Partially, the light is scattered by the metasurface and the handedness of the polarization is transformed, i.e. RCP light (as already explained in Supplementary Note 2), and the other component describes the incident LCP light which has not interacted with the metasurface. The incident LCP light is then filtered out with a second QWP and a rotated linear polarizer (LP).

For the beam steering metasurfaces (Figure 1 in main text), the detector is moved along a semicircle to detect the deflected RCP light. A QWP and a linear polarizer (LP) enables a clear distinction between both polarization chiralities (c.f. Figure 10A). In another experiment, the intensity profile of the metalens (Figure 2 in main text) is investigated. Therefore, the knife-edge method is used to determine the beam intensity profile at different positions behind the metalens measured with power sensor after filtering the incident LCP light (c.f Figure S10B). Accordingly, a razor blade is mounted at a micrometer stage and the varied laser power dependent on the position of the razor blade within the beam is recorded. The discrete derivative of the recorded power yields the lateral beam profile. The hologram and the OAM metasurfaces (Figure 3 and 4 in main text) are directly imaged with a thermal camera (TC) on a screen (c.f. Figure 10C and D). The spiral intensity pattern of the OAM metasurface caused by the direct interference with the incident light is obtained by slightly rotating the linear polarizer in order to achieve approximately similar laser powers of the converted RCP and incident LCP light. Finally, the dual-hologram (Figure 6 in main text) is directly imaged with a laser profiling camera Pyrocam IV due to increased sensitivity and more camera pixels to resolve the fine details of the targeted structure (c.f. Figure 10E).

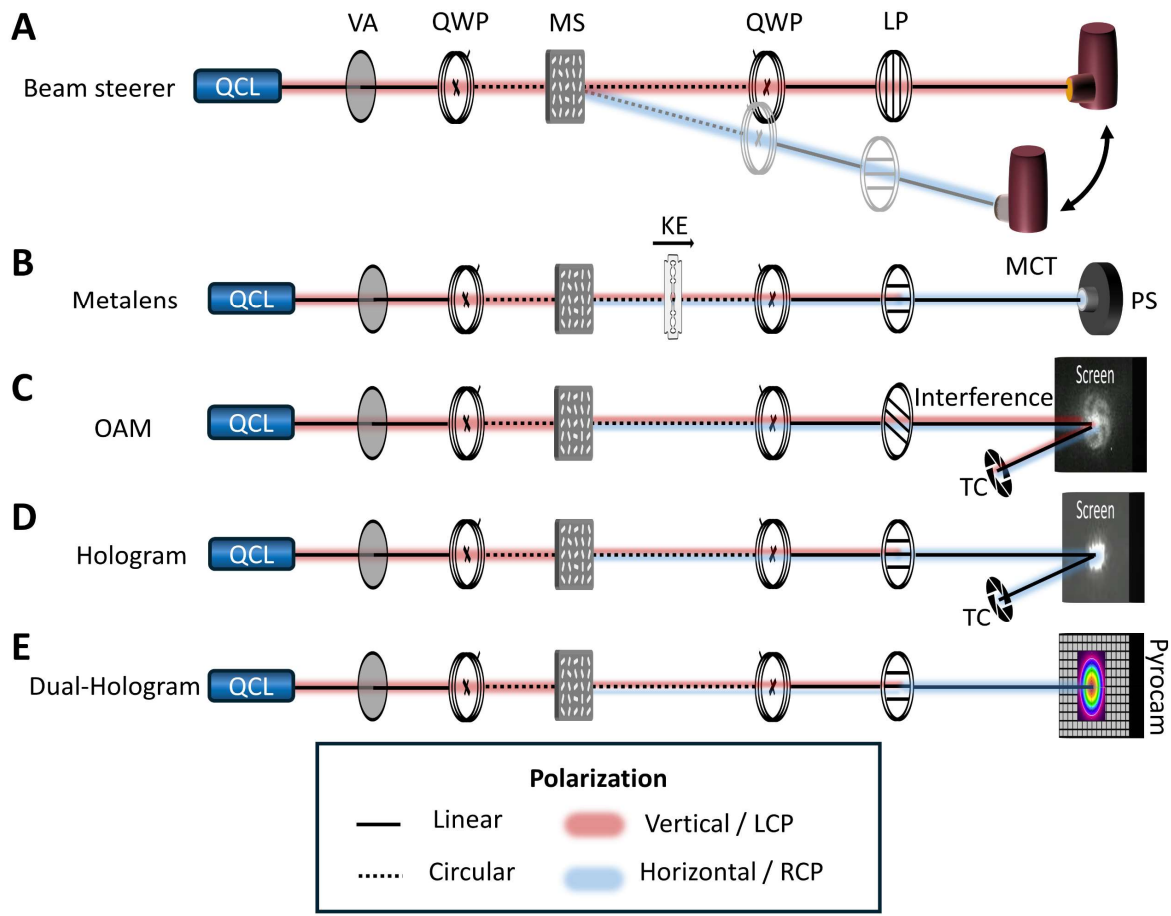

**Figure S10: Measurement Setups. A)** Characterization of the beam steering metasurfaces with angle-resolved measurements by rotating a detector on a semicircle. **B)** Characterization of the metalens by moving a knife-edge razor blade through the beam path and detecting the intensity with a power sensor (PS). **C)** The OAM metasurfaces are imaged at a screen with a conventional thermal camera (TC). Direct interference is obtained by rotating the linear polarizer (LP) to create equal intensity distributions of the incident light and the scattered light by the metasurface. **D)** The hologram is imaged with the TC on a screen. **E)** For improved accuracy and more detection pixels, the dual-hologram is directly imaged with a Pyrocam after filtering the incident LCP polarization at various distances behind the metasurface.